\documentclass[numsec,webpdf,modern,medium,namedate]{oup-authoring-template}
\onecolumn

\usepackage{bm}
\usepackage{multirow}
\usepackage{amsmath,amssymb}
\usepackage{booktabs}
\usepackage{lscape}
\usepackage{url}
\graphicspath{{Fig/}{figure/}}

\newcounter{subfigctr}
\AddToHook{env/figure/begin}{\setcounter{subfigctr}{0}}

\theoremstyle{thmstyleone}
\newtheorem{theorem}{Theorem}

\theoremstyle{thmstyletwo}

\theoremstyle{thmstylethree}

\definecolor{jade}{rgb}{0.0, 0.66, 0.42}

\begin{document}

\journaltitle{Journal of the Royal Statistical Society Series C: Applied Statistics}
\DOI{}
\copyrightyear{}
\pubyear{}
\access{}
\appnotes{Original article}
\firstpage{1}

\title[Varying-coefficient mixture of experts]{Varying-coefficient mixture of experts model for dynamic heterogeneous populations: application to mouse cortical development}
\author[1]{Qicheng Zhao}
\author[1,2,3]{Celia M.T. Greenwood}
\author[1,$\ast$]{Qihuang Zhang}
\authormark{Zhao et al.}
\address[1]{Department of Epidemiology, Biostatistics and Occupational Health, McGill University, Montréal, QC, Canada}
\address[2]{Gerald Bronfman Department of Oncology, McGill University, Montréal, QC, Canada}
\address[3]{Lady Davis Institute, Sir Mortimer B. Davis Jewish General Hospital, CIUSSS du Centre-Ouest-de-l'Île-de-Montréal, Montréal, QC, Canada}
\corresp[$\ast$]{Address for correspondence: Qihuang Zhang, Department of Epidemiology, Biostatistics and Occupational Health, McGill University, Montréal, QC, Canada. Email: \href{mailto:qihuang.zhang@mcgill.ca}{qihuang.zhang@mcgill.ca}}

\abstract{
As cells differentiate, gene--gene associations may change. Because the composition of cell subtypes also shifts with development, it is challenging to establish whether those changes reflect real changes in gene regulation within one or more subpopulations or a spurious one due to the shifting composition. Only formal inference can separate them. We propose a \emph{Varying-Coefficient Mixture-of-Experts} (VCMoE) model in which the coefficients of both the gating functions and the expert components vary smoothly with the index variable, so that a subgroup-specific association can be inferred without being confounded by the concurrent shift in subgroup composition. We develop a label-consistent EM algorithm and establish the identifiability and asymptotic theory that make the resulting inference valid. These guarantees support simultaneous confidence bands, constructed using asymptotic and bootstrap methods, together with a generalized likelihood ratio test for constant coefficient functions. Simulations demonstrate accurate estimation and satisfactory coverage. Applied to single-nucleus RNA-sequencing data from embryonic mouse cortex, VCMoE finds that the repression of \textit{Bcl11b} by \textit{Satb2} within the upper layer is not yet in place when those neurons first appear and becomes established over the course of differentiation, a dynamic that an ordinary mixture-of-experts model neglected. A package for implementing the method is available.}


\maketitle

\section{Introduction} 

How gene expression drives cell-fate specification, the process by which an uncommitted cell commits to a particular cell type, is a central question in developmental biology. Single-cell RNA sequencing (scRNA-seq) now makes this process observable at the resolution of individual cells in the developing embryo \citep{fei2022systematic}. Gene--gene associations, meaning how the expression level of one gene co-varies with another across cells, are of particular interest because they can reveal how developmental regulators interact with the marker genes that identify emerging cell types \citep{aibar2017scenic}. Prenatal mouse cortical development provides a case in point: as progenitor cells differentiate into distinct neuronal subtypes, gene--gene associations are expected to change, while the relative proportions of these subtypes also shift substantially over developmental time.

Therefore, the question we address is whether a gene--gene association genuinely varies over developmental time within one or more cellular states, rather than only appearing to do so because the latent cellular composition is shifting over the same period. Recently, the release of a single-cell time-lapse atlas of mouse prenatal development, that profiles 12.4~million nuclei from 83~embryos sampled at 2 to 6 hour intervals between gastrulation and birth \citep{qiu2024single}, provides rich data for addressing these questions, but suitable statistical tools remain limited. In current practice, cell states are first inferred by clustering and the estimated labels are then treated as known, which understates uncertainty in any subsequent association analysis \citep{traag2019louvain}. The deeper gap, however, is inferential: to our knowledge no existing method can accommodate a subgroup composition that shifts over developmental time while also testing whether a subgroup-specific association is significant and whether it genuinely varies rather than staying constant.

Although motivated by developmental scRNA-seq data, this challenge is a special case of a broader statistical problem involving dynamic heterogeneous populations. In many scientific settings, the objective is to characterize how covariate--outcome associations evolve along a meaningful external index, such as time, developmental stage, disease progression, dose, or spatial location \citep{zhang2023leveraging,liu2024single,srivatsan2020massively,zhao2026detecting}. At the same time, observations may arise from latent subgroups whose relative proportions also vary with the index. Consequently, changes observed at the population level may reflect either shifts in subgroup composition or genuine variation in subgroup-specific relationships. A unified framework is therefore needed to disentangle these sources of variation and provide valid inference for the underlying dynamic associations.

The mixture-of-experts (MoE) model provides a natural starting point for this problem. MoE is a \emph{conditional mixture} framework in which the conditional distribution of a response, given covariates, is represented as a weighted combination of expert regression models, with the mixing weights specified through covariate-dependent gating functions. In the context of latent heterogeneous populations, these gating weights can be interpreted as the probabilities that observations arise from different latent subgroups, while the corresponding expert models characterize subgroup-specific covariate--outcome associations. Thus, MoE provides an interpretable framework for jointly modeling latent subgroup membership and heterogeneity in covariate--outcome relationships.

The MoE framework was originally introduced by \citet{jacobs1991adaptive} in the context of neural network architectures and has since been extensively developed in the statistical literature \citep{grun2008flexmix}. Classical statistical MoE formulations typically specify the expert components as generalized linear models, such as Poisson, Gamma, or multinomial regressions, with coefficients that are fixed with respect to the indexing variable \citep{grun2008flexmix, jiang1999hierarchical, chen1999improved}. Consequently, although classical MoE models can accommodate latent heterogeneity, they are not designed to capture subgroup composition or subgroup-specific covariate--outcome associations that evolve smoothly over time, space, or another external index. Also, varying-coefficient models \citep{fan2008statistical} have been incorporated into finite mixture models by \citet{huang2018statistical}, but such models do not include the covariate-dependent gating mechanism that is central to MoE architectures. These limitations naturally motivate recent efforts to allow MoE components to vary with an external index.

Recent work has begun to develop dynamic MoE formulations. For example, \citet{chamroukhi2024functional} introduced B-spline-based varying coefficients within an MoE framework for scalar responses, and \citet{tamo2024mixture} extended this idea to functional responses. However, in the former, the coefficient functions weight the functional predictors rather than model covariate effects that evolve along an external index; in the latter, the gating function integrates the covariate trajectory over the whole index domain, so that the mixing proportions are constant over the index. Neither formulation therefore accommodates simultaneous changes
in subgroup-specific associations and subgroup composition along an index, as required in our motivating setting.

These approaches provide flexible tools for prediction in dynamic mixture models, but two features limit their use in our setting. They are developed primarily for Gaussian responses, whereas scRNA-seq gene-expression counts are overdispersed counts, and are commonly modeled using negative-binomial distributions \citep{zhang2019probabilistic}. More importantly, the objective here is inference rather than prediction: to determine whether subgroup-specific gene--gene associations are statistically significant and whether they genuinely vary over developmental time, for which these methods provide no uncertainty quantification or formal tests.

Another related direction is the dynamic MoE model of \citet{munezero2023dynamic}, which allows the parameters in both the expert components and the gating functions to evolve over time through random-walk state processes. This framework is designed for online prediction, in the sense that the predictive distribution is sequentially updated as new observations become available. Although well suited to real-time forecasting, this formulation represents temporal variation through a sequence of time-specific latent states rather than through smooth coefficient functions defined over a continuous indexing domain. Consequently, it does not directly provide inferential tools for assessing whether a covariate effect is statistically significant, remains constant, or varies smoothly along the index. Such inference is exactly what our motivating developmental scRNA-seq problem requires, where the objective is to characterize dynamic gene--gene associations within latent cellular states while accounting for changes in subgroup composition.

To address these limitations, we propose the \emph{Varying-Coefficient Mixture-of-Experts} (VCMoE) model, which allows the coefficients in both the gating functions and the expert components to vary smoothly with an indexing variable. We also provide an accompanying R package, \texttt{VCMoE}, available from CRAN.

In this article, we make four major theoretical and computational developments for the proposed VCMoE framework designed to characterize the dynamic relationships between exposures and outcomes under the evolving subpopulations:
\begin{enumerate} \item 
The identifiability and consistency of the VCMoE model are rigorously examined under regularity conditions.  
\item A tailored expectation-maximization (EM) algorithm is proposed to estimate the functional coefficients. This procedure accommodates both fully functional (i.e., all coefficients vary) and hybrid specifications (i.e., only a subset of coefficients varies). The asymptotic distributions of the resulting estimators are also derived.  
\item Simultaneous confidence bands are constructed using both asymptotic theory and a parametric bootstrap approach. ]item Three hypothesis testing procedures, including asymptotic, bootstrap-based, and generalized likelihood ratio tests, are introduced to statistically assess whether specific coefficients are varying or not.
\end{enumerate}

The remainder of the paper is organized as follows. Section~\ref{sec_main_model} introduces the proposed model formulation and presents theoretical results establishing identifiability. In Section~\ref{sec_global_local_est}, a label-consistent EM algorithm is developed for parameter estimation, and the asymptotic properties of the resulting estimators are derived. Section~\ref{sec_cb} details the construction of simultaneous confidence bands and outlines associated hypothesis testing procedures. Section~\ref{sec_sim} reports the results of simulation studies conducted across a range of settings. Finally, Section~\ref{sec:Application} demonstrates the utility of the proposed methodology through its application to a dataset of single-nucleus RNA sequencing (snRNA-seq) gene expression obtained from embryonic mice between embryonic day~14 and embryonic day~18.75.

\section{Varying-coefficient Mixture of Experts Model} \label{sec_main_model}

For $i=1, \dots, n$, let \( Y_i \) denote the outcome of subject $i$, drawn from a population composed of \( C \) latent subpopulations; in our motivating application, \(Y_i\) is the measured expression of one gene in cell $i$ and the \(C\) subpopulations are distinct neuronal cell states. The membership of each observation to a specific subpopulation is unobserved and represented by a latent categorical variable \(\mathcal{C}_i\). Let $\boldsymbol{x}_i$ and $\boldsymbol{z}_i$ denote the covariate vectors associated with observation $i$, where \(\boldsymbol{z}_i\) carries the expression of a second gene whose association with \(Y_i\) is of scientific interest and \(\boldsymbol{x}_i\) collects covariates that inform subpopulation membership. Furthermore, let $U$ represent a continuous index variable, such as developmental time or a one-dimensional spatial location, at which the response $Y_i$ is observed. Conditional on this scalar index variable \( U \) and $\boldsymbol{x}_i$, the probability that \(i\) is allocated to subpopulation $c$ is assumed to be 
\( P(\mathcal{C}_i = c \mid u, \boldsymbol{x}_i) = \pi_{c}( \boldsymbol{x}_i; \boldsymbol{\beta}_c(u)), \ \text{for} \; c = 1, \ldots, C \). In most mixture-of-experts frameworks, the component probabilities 
\(\pi_{c}(\cdot)\) are typically specified as functions of the covariate vector \(\boldsymbol{x}_i\), with coefficients 
\(\boldsymbol{\beta}_c\). In our formulation, we extend this by 
allowing the coefficient vector \(\boldsymbol{\beta}_c\) to vary with \(U\), 
yielding the form \(\pi_{c}( \boldsymbol{x}_i; \boldsymbol{\beta}_c(u))=g(\boldsymbol{x}_i^\top \boldsymbol{\beta}_c(u))\). 
The function \(g(\cdot)\) is commonly referred to as \textit{gating function}.
 For a given value \( U = u \) and corresponding covariate vector $\boldsymbol{x}_i$, the probabilities naturally satisfy that
\( \pi_{1}( \boldsymbol{x}_i; \boldsymbol{\beta}_1(u)) + \cdots + \pi_{C}( \boldsymbol{x}_i; \boldsymbol{\beta}_C(u)) = 1 \), for each $i \in \left\{1, \dots, n\right\}$.

For each subpopulation, the conditional distribution of \(Y_i\) given 
\(\boldsymbol{z}_i\) may differ. Specifically, we assume that within 
subpopulation \(c\), the expert model follows a distribution with 
density function \(\phi(\cdot)\), parameterized 
by the mean \(\eta_c(\boldsymbol{z}_i;\boldsymbol{\alpha}_c(u))\) and the dispersion parameter \(\delta_c(u)\). Without knowledge of the specific subpopulation to which subject \(i\) belongs, 
the conditional density of \(Y_i\), given \(U = u\), 
can be expressed as
\begin{equation}
    \sum_{c=1}^C g\!\left( { \boldsymbol{x}_i^{\top} \boldsymbol{\beta}}_c(u) \right) \,
    \phi \!\left\{ y_i \,\middle|\, \eta_c(\boldsymbol{z}_i;\boldsymbol{\alpha}_c(u)), \, \delta_c(u) \right\},
    \label{main_model}
\end{equation}
where \(\eta_c(\boldsymbol{z}_i;\boldsymbol{\alpha}_c(u)) = w( \boldsymbol{z}_i^\top \boldsymbol{\alpha}_c(u))\) denotes the conditional mean function. Here, $\phi(\cdot)$ is known as the expert model and $w(\cdot)$ is an inverse link function. As an illustration, the density function $\phi(\cdot)$ is taken to be a member of the general exponential family, although the framework extends beyond this case.

We note that, in our model formulation, while different notations are used to denote the covariates in the gating function and the expert models, these covariates may or may not represent the same variables, unlike the conventional MoE framework where two covariates are commonly assumed to be identical. We distinguish them here to provide greater generality beyond the standard MoE setup. If the covariates overlap partially or completely, the identifiability of the parameters becomes an important consideration addressed in Supplementary Material S1.

\section{Estimation}\label{sec_global_local_est}

Our estimation strategy proceeds in stages. We first approximate each coefficient function locally and estimate it by maximizing a kernel-weighted local likelihood (Section~\ref{subsec_est}), using a label-consistent EM algorithm to keep the subpopulation labels coherent across the index. Coefficients that are in fact constant are then estimated by a two-step refinement that averages the local fits to recover the faster parametric rate (Section~\ref{est_cons}). We select the bandwidth by likelihood cross-validation (Section~\ref{sec:bandwidth_selection}) and establish the asymptotic distributions of the resulting estimators, including the maximum deviation that underpins the simultaneous confidence bands (Section~\ref{asy_prop}).

\subsection{Local estimator}\label{subsec_est}
The consistency of a global estimator for Model (\ref{main_model}) relies on a rather restrictive assumption, namely compactness (see detailed discussion in Supplementary Material S2). In this section, we address this restriction by an alternative approach to global estimation, local regression. The local regression employs a Taylor expansion to construct a local estimator, thereby allowing flexibility not accessible to the global estimator.

For a fixed \(u\), the local model can be expressed as a weighted likelihood of a finite mixture model, and the local estimators of $\boldsymbol{\alpha}(u), \boldsymbol{\beta}(u), \text{and } \delta(u)$ are the maximizers of the following local log-likelihood function,
\begin{equation}\label{log_local_likelihood}
    \ell_{n} = \frac{1}{n}\sum_{i=1}^n \log \left(    \sum_{c=1}^C \pi_{c}(\boldsymbol{x}_i;\boldsymbol{\beta}_c(u)) \,
    \phi \!\left\{ Y_i \,\middle|\, \eta_c(\boldsymbol{z}_i;\boldsymbol{\alpha}_c(u)), \, \delta_c(u) \right\}
    \right)K_h(U_i-u),
\end{equation}
where $K_h(t) = K(t/h)/h$, with $K(t)$ denoting a kernel function and $h$ representing a prespecified bandwidth. The resulting estimator $\hat{\boldsymbol{{\theta}}}(u)=\left\{\hat{\boldsymbol{\alpha}}(u), \hat{\boldsymbol{\beta}}(u),\hat{\delta}(u)\right\}$ is obtained by maximizing the local log-likelihood function (\ref{log_local_likelihood}). 

For clarity of exposition, we restrict our attention to a two-component mixture model for the remainder of the presentation of the theory. In particular, the mixing proportions for observation $i$ are modeled as
$
\pi_{1}(\boldsymbol{x}_i;\boldsymbol{\beta}(u)) = \operatorname{expit}\!\bigl(\boldsymbol{x}_i^\top \boldsymbol{\beta}(u_i)\bigr), 
\ \text{and}\
\pi_{2}(\boldsymbol{x}_i;\boldsymbol{\beta}(u)) = 1 - \operatorname{expit}\!\bigl(\boldsymbol{x}_i^\top \boldsymbol{\beta}(u_i)\bigr),
$
where \(\operatorname{expit}(x) = \frac{1}{1 + e^{-x}}\). 
The proposed methodology can be readily extended to mixtures with \(C > 2\) components 
by adopting suitable link functions; for instance, the softmax function is a common choice 
when the number of classes is three or more.

In local regression, an essential consideration concerns the order of approximation applied to the coefficient functions, e.g., $\boldsymbol{\beta}(u_i)$. Possible choices include local constant, local linear, or higher-order polynomial approximations. In this work, we adopt the local linear approximation for each coefficient function, as the local linear framework has been shown to have several appealing advantages, such as statistical efficiency, adaptability to the design, and favorable boundary behavior \citep{fan1993local, ruppert1994multivariate}.
 Specifically, assume that \( \beta_p(U_i) \), which is a $p$-th element in the $\boldsymbol{\beta}$, possesses a continuous second derivative. For any given \( u \), applying a Taylor expansion yields
\begin{equation}
\begin{aligned}
     \beta_p(U_i) &\approx \beta_p(u) + h \beta_p'(u) \frac{U_i - u}{h}.
\end{aligned}\label{local_linear_expansion}
\end{equation}
 This indicates that under a local linear expansion, the coefficient functions can be approximated by the addition of the function value at \(u\) 
and the local slope (i.e., the first derivative) of the function evaluated at \(u\).  The similar local linear approximation can be applied to $\alpha_p(U_i)$ and \(\delta(U_i)\).

\subsubsection{Label-consistent EM algorithm}\label{m_em}
In practice, the Expectation--Maximization (EM) algorithm serves as a natural estimation approach for estimating $\hat{\boldsymbol{\theta}}(u)$. 
However, a purely pointwise implementation, where the component labels are treated independently across local models 
at each specific \(u\), poses challenges due to label switching. 
When the model is fitted independently at each \(u\), the resulting component labels fail to remain consistent across neighboring locations. To resolve this, we propose a label-consistent EM algorithm \citep{huang2013nonparametric} for parameter estimation in the model. 

To estimate coefficient functions at each given point $u$, we employ a modified EM algorithm in which the E-step estimates component memberships globally, independent of the specific location \(u\), while in the M-step, the component-specific coefficient functions are updated simultaneously over a set of grid points, $\left\{u: u \in [0,1]\right\}$. This step ensures consistent labeling and smooth functional estimation.  Based on this representation, the modified EM algorithm proceeds with iterating the following steps.

\textbf{E-step}: In iteration $t$, for $i=1,\dots,n$, with a given $\boldsymbol{\theta}_c^{t-1}(u_i)=\left\{\boldsymbol{\beta}^{t-1}_c(u_i), \boldsymbol{\alpha}^{t-1}_c(u_i),\delta^{t-1}_c(u_i)\right\}$, for $c \in \left\{1,2\right\}$, we calculate 
$
    \gamma_{ic}=\frac{\pi_{c}(u_i;\boldsymbol{x_i},\boldsymbol{\beta}_c^{t-1}(u_i)) \phi(y_i \mid \eta_c(\boldsymbol{z}_i;\boldsymbol{\alpha}^{t-1}_c(u)), \delta_c^{t-1}(u_i))}{\sum_{c=1}^2 \pi_{c}(u_i;\boldsymbol{x_i}, \boldsymbol{\beta}_c^{t-1}(u_i)) \phi(y_i \mid \eta_c(\boldsymbol{z}_i;\boldsymbol{\alpha}^{t-1}_c(u)), \delta_c^{t-1}(u_i))},
$
where $\pi_{c}(\cdot)$ and $\phi(\cdot)$ retain the same definitions as provided in Section~\ref{subsec_est}.

\textbf{M-step}: Given $\boldsymbol{\gamma}_c=(\gamma_{1c}, \dots, \gamma_{nc})$, for a fixed grid point $u \in \mathcal{U}$, we update $\boldsymbol{\theta}_c(u)$ by maximizing the following function with respect to $\boldsymbol{\theta}_c(u)=\left\{\boldsymbol{\beta}_c(u), \boldsymbol{\alpha}_c(u),\delta_c(u)\right\}$ taking the local linear expansion as in (\ref{local_linear_expansion}),
$
     \boldsymbol{Q}(\boldsymbol{\theta}_c(u)|\gamma_{ic})=\sum_i^n \left\{\sum_{c=1}^2 \gamma_{ic}\log\left\{\phi(y_i \mid \eta_c(\boldsymbol{z}_i;\boldsymbol{\alpha}_c(u)), \delta_c(u))\right\}  K_h(U_i-u)   \right\}+\sum_i^n \left\{\sum_{c=1}^2\gamma_{ic} \log(\pi_c) K_h(U_i-u)\right\}.
$

\subsubsection{Estimation of constant coefficient}\label{est_cons}

The estimation framework presented in Section~\ref{m_em} builds on the premise that the coefficients are functions rather than constants, and it is therefore inefficient to directly apply such an estimation procedure to a constant coefficient setting. This oversight can induce an inflated variance in the estimator that is mistakenly regarded as varying, thereby reducing power to detect the covariate effect. In this section, we propose an estimation framework for a coefficient under the null assumption that it remains constant.

Suppose that one specific coefficient function, \(\beta_j(\cdot)\), is in fact constant, denoted by \(\beta_j\). The subscript \(c\) is omitted since, in the two-class model, only a single coefficient vector \(\boldsymbol{\beta}\) is required. We propose a two-step estimation procedure for \(\beta_j\), following an idea originating in \citet{zhang2002local} for a simpler setting. In Step 1, \(\beta_j\) is estimated as though it were a function, following the procedure of Section~\ref{m_em}. In Step 2, the constant coefficient is obtained by averaging the local estimates, that is, for $j \in \{1, \dots, p_\beta\}$, where $p_\beta$ denotes the dimension of $\boldsymbol{\beta}$,
\begin{equation}\label{eqn_cons_cal}
\hat{\beta}_{j} = \frac{1}{n}\sum_{i=1}^n \hat{\beta}_{j}(u_i).
\end{equation}

The intuition is as follows, in Step 1, treating $\beta_{j}(\cdot)$ as a function produces an estimator with relatively large variance, while in Step 2, averaging across locations reduces this variance. The same strategy applies to the estimation of $\alpha_{cj}$ and $\delta_{c}$. This two-step procedure can be seamlessly incorporated into the \textbf{M-step} of the modified EM algorithm introduced in Section~\ref{m_em}, requiring only the substitution of $\hat{\beta}_{j}$ with the expression in (\ref{eqn_cons_cal}) after each iteration. In Section~\ref{asy_prop}, we show that the resulting estimator is asymptotically normal with convergence rate $O_p(n^{-1/2})$, provided the bandwidth is selected within a suitable range. Since the convergence rate for the constant coefficient estimator is $O_p(n^{-1/2})$, the estimation of the remaining functional coefficients attains the same asymptotic properties as if $\beta_j$ were known, due to their convergence rate of order $(nh)^{1/2}$.

\subsection{Bandwidth Selection}\label{sec:bandwidth_selection} 
Bandwidth selection is a key issue in kernel-based nonparametric modeling. 
In this paper, we adopt the likelihood cross-validation (CV) approach discussed in \cite{zhang2010simultaneous}. 
Specifically, for each $i = 1, \dots, n$, we omit the $i$th observation and estimate $\boldsymbol{\theta}(u_{i,h})$ using the remaining data with bandwidth $h$. 
The resulting estimator is denoted by $\hat{\boldsymbol{\theta}}^{\setminus i}(u_{i,h})$. 
This yields the following leave-one-out cross-validation criterion:
$
\text{CV}(h) = \sum_{i=1}^{n} \log \left(    \sum_{c=1}^C \pi_{c}(\boldsymbol{x}_i;\hat{\boldsymbol{\beta}}^{\setminus i}_c(u_{i,h})) \,
    \phi \!\left\{ Y_i \,\middle|\,\eta_c(\boldsymbol{z}_i;\boldsymbol{\alpha}^{\setminus i}_c(u_{i,h})), \, \hat{{\delta}}^{\setminus i}_c(u_{i,h}) \right\}
    \right).
$ The optimal bandwidth is then chosen as the value of $h$ that maximizes $\text{CV}(h)$.

\subsection{Asymptotic properties}\label{asy_prop}
In this section, we establish the asymptotic properties of the local coefficient estimators, $\hat{\boldsymbol{\theta}}(u)$, described in Section \ref{m_em}. To ease the notation, let $f(y_i \mid \boldsymbol{x}_i, \boldsymbol{z}_i,\boldsymbol{\theta}(u))
= \sum_{c=1}^{2} \pi_{c}(\boldsymbol{x}_i;\boldsymbol{\beta}(u)) \,\phi\!\left\{\, y_i \mid \eta_c(\boldsymbol{z}_i;\boldsymbol{\alpha}_c(u)),\ \delta_c(u) \right\}$  denote the conditional density defined in (\ref{main_model}), with the formulation restricted to the two-class case. Then, we denote $\ell(\boldsymbol{\theta}(u);\boldsymbol{x}_i,\boldsymbol{z}_i,y_i) = \log f(y_i \mid \boldsymbol{x}_i, \boldsymbol{z}_i,\boldsymbol{\theta}(u)),$ and $q_{\boldsymbol{\theta \theta}}(\boldsymbol{\theta}(u);\boldsymbol{x}_i, \boldsymbol{z}_i ,y_i)
= \frac{\partial^{2} \ell(\boldsymbol{\theta}(u);\boldsymbol{x}_i,\boldsymbol{z}_i,y_i)}{\partial \boldsymbol\theta\,\partial \boldsymbol{\theta}^{\top}}$. We impose regularity conditions (RC 1)--(RC 9), which are listed in Supplementary Section~S9. We first establish Lemma~1 (see Supplementary Material~S9), which proves the consistency of the maximum likelihood estimator (MLE). Building on this result, we present the following theorem on its asymptotic properties.
\begin{theorem}\label{main_asy}
    Assume the regularity conditions (RC 1)-(RC 9) hold. Then, with probability approaching to 1, there exists a consistent local maximizer, $\boldsymbol{\hat{\theta}}(u)$ satisfy the following 
\begin{align*}
\sqrt{nh}\Big\{\hat{\boldsymbol{\theta}}(u) - \boldsymbol{\theta}(u) 
-  \left[ \frac{h^2}{2}\boldsymbol{\theta}''(u)v_2+o_p(h^2) \right] \Big\} \xrightarrow{D} \mathcal{N}\Bigg(\mathbf{0}_{p_{\theta}},\, \tau f^{-1}(u)\,\mathcal{I}^{-1}(u)\Bigg),
\end{align*}
    where $p_{\theta}$ is the dimensionality of $\boldsymbol{\theta}$, $\mathbf{0}_{p_{\theta}}$ is a $p_{\theta} \times1$ vector with each entry being 0, $\tau=\int K^2(u)du$, and $v_2=\int u^2 K(u)du$.
\end{theorem}

 Following Theorem~\ref{main_asy}, the asymptotic bias of the estimator 
$\hat{\boldsymbol{\theta}}$ is given by  
\begin{equation}\label{asy_bias}
\frac{h^2}{2}\boldsymbol{\theta}''(u)v_2\{1+o_p(1)\}.    
\end{equation}

 As it plays a pivotal role in constructing simultaneous confidence bands and conducting hypothesis testing within the 
varying-coefficient model framework, we discuss its estimation here. Following (\ref{asy_bias}) and in line with the approach of \cite{zhang2010simultaneous}, we propose the following estimator of the bias of $\hat{\boldsymbol{\theta}}(u)$,
\begin{align} \label{est_bias}
    \widehat{\operatorname{bias}}(\hat{\boldsymbol{\theta}}(u)\mid\mathcal{D})=\frac{h^2}{2} \hat{\boldsymbol{\theta}}''(u)v_2.
\end{align}
Here, the estimator $\hat{\boldsymbol{\theta}}''(u)$ of $\boldsymbol{\theta}''(u)$ can be obtained by local cubic maximum likelihood estimation with an appropriate pilot bandwidth, which may be chosen according to the method of \cite{fan1996study}. In practice, however, it is often difficult to accurately estimate the bias of $\hat{\boldsymbol{\theta}}(u)$ due to the instability of higher-order derivatives estimation. Consequently, bias estimation via (\ref{est_bias}) is primarily for theoretical discussion \citep{zhang2010simultaneous}. A practical alternative is to use a smaller bandwidth so that the bias becomes negligible.

Another important component when constructing confidence bands or carrying out hypothesis tests is the estimation of variance. We use a sandwich estimator of the covariance matrix, with implementation details provided in Supplementary Material~S3.

Next, we study the asymptotic distribution of the maximum discrepancy between the estimated functional coefficient and its true counterpart. This result forms the basis for constructing simultaneous confidence bands and for the hypothesis testing procedure discussed later. To our knowledge, this is the first time simultaneous confidence bands have been extended to the mixture model. 

Without loss of generality, we assume that the domain of $\mathcal{U}$ is $[0,1]$, since the support set can typically be standardized to this scale. Let $\widehat{\operatorname{bias}}(\hat{\beta}_p(u)\mid \mathcal{D})$ denote the $p$th component of $\widehat{\operatorname{bias}}(\hat{\boldsymbol{\beta}}(u)\mid \mathcal{D})$, and let $\widehat{\operatorname{var}}(\hat{\beta}_p(u)\mid \mathcal{D})$ denote the $p$th diagonal element of $\widehat{\operatorname{Cov}}(\hat{\boldsymbol{\beta}}(u)\mid \mathcal{D})$. The same result holds for $\hat{\boldsymbol{\alpha}}(u)$ and $\hat{\delta}(u)$.

\begin{theorem}\label{th_max_dis}
Under regularity conditions (RC 1)–(RC 9), together with the assumptions stated in Lemma~A.2 of the Supplementary Material, 
and for a bandwidth \(h = O(n^{-b})\) with \(1/5 \leq b < 1 - 2/s\), 
where \(s\) denotes the moment-order parameter as defined in Lemma~A.2, we have for any $r \in \mathbb{R}$
{\footnotesize
\begin{align*}
P \left\{ 
(-2 \operatorname{log} h)^{1/2} 
\left( \sup_{u \in [0,1]}\Bigg| 
\frac{1}{\widehat{\operatorname{var}}(\hat{\beta}_p(u)\mid \mathcal{D})^{\frac{1}{2}}}
\left( \hat{\beta}_p(u) - \beta_p(u) - \widehat{\operatorname{bias}}(\hat{\beta}_p(u)\mid \mathcal{D}) \right) 
\Bigg|- d_{\nu,n} \right) < r
\right\}
\;\;\longrightarrow\;\; \exp\{-2 \exp(-r)\},
\end{align*}}
\end{theorem}
where $d_{v,n}$ corresponds to $d_n$, which is defined as
$d_n = (-2\log h)^{1/2}
\allowbreak + \frac{1}{(-2\log h)^{1/2}}
\allowbreak \left\{ \log \frac{K^2(A)}{\nu_0 \pi^{1/2}}
\allowbreak + \tfrac{1}{2} \log \!\log h^{-1} \right\}$
or
$d_n = (-2\log h)^{1/2}
+ \frac{1}{(-2\log h)^{1/2}}
\log \left\{ \frac{1}{4\nu_0\pi} \int (K'(t))^2 dt \right\}$ under different choices of the kernel function, as discussed in Lemma A.2 of the Supplementary Material; here $v_0$ and $K(u)$ are replaced by $v_{1,0}$ and $K_1(u)$, respectively.

Next, we study the asymptotic properties of the two-step estimator for the constant coefficient, showing that its convergence rate is $O_p(n^{-1/2})$. It should be noted that this convergence rate is substantially faster than that of the functional coefficient estimator.

\begin{theorem}\label{theorem_con_est}
Under the regularity conditions (RC 1)-(RC 9), when $\beta_p(u)$ is a constant $\beta_p$, if $h \to 0$, $\sqrt{n}h^2 \to 0$ and $nh^2/(-\operatorname{log} h) \to \infty$, then
\[
\sqrt{n}(\hat{\beta}_p - \beta_p-O_p(h^2)) \;\; \xrightarrow{D} \;\; \mathcal{N}(0, \sigma_c^2),
\]
where $e_{p,p}$ denotes a \(p\)-dimensional unit vector whose \(p\)th element equals to one and all other elements are zero, $\sigma_c^2=\mathbb{E}\big(e^\top_{p,p}\mathcal{I}^{-1}(U) e_{p,p}\big)$.
\end{theorem}

From Theorem~\ref{theorem_con_est}, we note that convergence to a non-degenerate limit implies tightness. Consequently, we have $\sqrt{n}(\hat{\beta}_p - \beta_p - O_p(h^2)) = O_p(1)$. Moreover, since $\sqrt{n}h^2 \to 0$, the bias term becomes negligible, and we can therefore conclude that the convergence rate is $O_p(n^{-1/2})$.

Then, building upon Theorem \ref{th_max_dis} and Theorem \ref{theorem_con_est}, if $\beta_p$ is in fact a constant, we have the following result about the asymptotic distribution of the maximum discrepancy, which provides a convenient basis for hypothesis testing:

\begin{theorem}\label{theorem_maxi_constant}
    Under the same conditions as in Theorem \ref{th_max_dis} and Theorem \ref{theorem_con_est}, we have for any $r \in \mathbb{R}$,
{\footnotesize
\begin{equation*}
P\left\{ 
(-2 \log h)^{1/2} 
\left( 
\sup_{u \in [0,1]} \Bigg|
\frac{1}{\{\widehat{\operatorname{var}}(\beta_p(u)\mid \mathcal{D})\}^{1/2}} 
\left( \hat{\beta}_p(u) - \hat{\beta}_p - \widehat{\operatorname{bias}}(\beta_p(u)\mid \mathcal{D}) \right) \Bigg|
- d_{\nu,n} 
\right) < r 
\right\} 
\;\;\longrightarrow\;\; \exp\{-2 \exp(-r)\}.
\end{equation*}}
\end{theorem}

Theorem~\ref{theorem_maxi_constant} extends Theorem~\ref{th_max_dis} 
to the setting where the true coefficient $\beta_p$ is constant rather than a function,  a case that, to our knowledge, has not been previously studied. Consequently, this theorem provides a foundational framework for testing whether the coefficient 
varies with $u$ or remains constant, as further discussed in Section~\ref{subsec_ht}.

\section{Confidence bands and hypothesis tests}\label{sec_cb}
\subsection{Confidence bands}\label{subsec_cb}
For nonparametric modeling, a pointwise confidence band quantifies uncertainty only at a single position $u_i$, whereas a simultaneous confidence band quantifies it across the entire coefficient function. We focus on the latter.
The construction of such bands relies on the distribution of the maximum discrepancy between the true coefficient function and the estimated coefficient function. In this section, we present two ways in addressing maximum discrepancy: an asymptotic approach and a bootstrap approach. In the discussion here, without loss of generality, we assume that $\mathcal{U} = [0,1]$. If not, the time range can be scaled to satisfy this assumption.

\subsubsection{Asymptotic distribution-based approach}\label{cb_asy}
The construction of simultaneous confidence bands using the asymptotic distribution is relatively straightforward. Based on Theorem~\ref{th_max_dis}, the following $100(1-\rho)\%$ confidence band for $\beta_p$ over the interval $u \in [0,1]$ can be readily derived,
$
    \hat{\beta}_p(u) - \widehat{\operatorname{bias}}(\hat{\beta}_p(u) \mid \mathcal{D}) \pm \Delta_{\rho}(u),
$
for a bandwidth $h$, where 
$
\Delta_{\rho}(u) = \Bigl(d_{v,n} + \bigl[\log 2 - \log \{-\log(1-\rho)\}\bigr](-2 \log h)^{-1/2}\Bigr)\Bigl\{\widehat{\operatorname{var}}(\hat{\beta}_p(u)\mid \mathcal{D})\Bigr\}^{1/2}.
$
This confidence band guarantees that with probability $1 - \rho$, it covers the true $\beta_p(u)$ for all $u \in [0,1]$.

\subsubsection{Bootstrap-based approach}\label{cb_boot}

The asymptotic approach is primarily preferable in its ease of implementation and low computational cost. Nevertheless, when the sample size is limited, the coverage probability of the resulting confidence band may be unsatisfactory. The bootstrap approach provides an alternative method for constructing simultaneous confidence bands. Compared with the asymptotic approach, the bootstrap typically yields more reliable uncertainty quantification when the sample size is small to moderate. The trade-off, however, is that the bootstrap procedure requires substantially greater computational time.

We define
$
T_p=\sup_{u \in [0,1]} \frac{|\hat{\beta}_p(u)-\beta_p(u)|}{\{\operatorname{var}(\hat{\beta}_p(u)\mid\mathcal{D})\}^{1/2}},
$
where $T_p$ represents the maximum standardized deviation between 
the estimated function $\hat{\beta}_p(u)$ and the true function $\beta_p(u)$ 
across the entire domain $u \in [0,1]$. Suppose the upper $\rho$ quantile of the distribution of $T_p$ is $c_{\rho}$. If both $c_{\rho}$ and $\operatorname{var}(\hat{\beta}_p(u)\mid \mathcal{D})$ were known, the confidence band of $\beta_p(\cdot)$ on the interval $[0,1]$ can be constructed as
\begin{equation}\label{eqn_boootstrap}
\hat{\beta}_p(u) \pm \{\operatorname{var}(\hat{\beta}_p(u)\mid\mathcal{D})\}^{1/2} c_{\rho}.
\end{equation}

In practice, both $c_{\rho}$ and $\operatorname{var}(\hat{\beta}_p(u)\mid \mathcal{D})$ are unknown and can be estimated via bootstrap. Suppose we obtain the estimators $\hat{c}^*_{\rho}$ and $\widehat{\operatorname{var}}^*(\hat{\beta}_p(u)\mid \mathcal{D})$ for $c_{\rho}$ and $\operatorname{var}(\hat{\beta}_p(u)\mid \mathcal{D})$, respectively. Substituting these estimates into (\ref{eqn_boootstrap}) yields the $(1-\rho)$ simultaneous confidence band of $\beta_p(\cdot)$:
$
\hat{\beta}_p(u) \pm \{\widehat{\operatorname{var}}^*(\hat{\beta}_p(u)\mid \mathcal{D})\}^{1/2}\hat{c}^*_{\rho}.
$

The detailed steps for implementing the bootstrap procedure to estimate $c_{\rho}$ and $\operatorname{var}(\hat{\beta}_p(u)\mid \mathcal{D})$ are provided in Supplementary Material S4.

\subsection{Hypothesis tests for constant coefficients}\label{subsec_ht}
In the proposed model, all coefficients in the component models and mixing proportions are allowed to vary, so a central inferential question is whether a given coefficient function is in fact constant. In our application, this asks whether a gene--gene association genuinely varies over developmental time rather than remaining fixed. For the two-class case, without loss of generality, we consider the following hypothesis concerning the $p$th component of $\boldsymbol{\beta}(\cdot)$:
\begin{equation}\label{null_hyp}
    H_0:  \beta_p(\cdot) \ = \beta_p, \ \text{and} \ H_a: \beta_p(\cdot) \ \neq \beta_p.
\end{equation}
It is important to note that the null and alternative hypotheses stated above are nonparametric, and the numbers of parameters under $H_0$ and $H_a$ are not well defined. In this section, we discuss three approaches to hypothesis testing. The first approach relies on asymptotic distribution, the second one employs a bootstrap-based procedure, and the third is constructed using the generalized likelihood ratio test. The asymptotic and bootstrap approaches parallel the constructions in Section~\ref{subsec_cb}, so their details are deferred to Supplementary Material~S5; we present the generalized likelihood ratio test below.

\subsubsection{Generalized likelihood ratio approach}\label{glrt}
The generalized likelihood ratio test (GLRT) proposed by \cite{fan2001generalized} is a powerful method for hypothesis testing in nonparametric models. Let $\ell_n(H_0)$ and $\ell_n(H_a)$ denote the log-likelihood functions under the null and alternative hypotheses, respectively, and define the generalized likelihood ratio test statistic as
$
    \lambda_n=\ell_n(H_a)-\ell_n(H_0).
$

In the following theorem, we show that the generalized likelihood ratio statistic $\lambda_n$, with a suitably chosen normalization constant, follows an asymptotic chi-squared distribution, and thereby can establish a Wilks-type result.

\begin{theorem} \label{theorem_glrt}
    Suppose that the regularity conditions (RC 1)--(RC 9) hold and assume the support set of $u$ is $[0,1]$. Then, under $H_0$, as $h \rightarrow 0$, $nh^{3/2} \rightarrow \infty$ and $nh^{9/2} \rightarrow 0$, we would have $r_K \lambda_n \xrightarrow{D} \chi^2_\delta$,
    where $r_K=[K(0)-0.5\int K^2(u)du]/\int[K(u)-0.5K*K(u)]^2du$, $\delta=r_K p_{\beta} C [K(0)-0.5\int K^2(u) du]/h$, and $K * K(u)$ is the second convolution of $K(\cdot)$.
\end{theorem}

Here, $p_{\beta}$ is the dimension of $\boldsymbol{\beta}$ in the hypothesis and $C$ is the number of classes. Hence, $p_{\beta}C$ is given by the total number of parameters under test, and can be easily adjusted to the specific null hypothesis under different considerations.

\section{Simulation studies} \label{sec_sim}
In this section, we conduct simulation studies under four scenarios to evaluate the performance of the proposed model: 
(i) a two-component Gaussian expert model, 
(ii) a two-component binomial expert model, 
(iii) a three-component Gaussian expert model, 
(iv) five-component models evaluated separately under Gaussian, binomial, and negative-binomial responses, all with a common gating structure. 
The first two scenarios assess the applicability of the proposed approach to continuous and discrete responses, respectively. 
The third and fourth scenarios examine its extension to mixtures with more than two components, for which the gating function adopts a softmax form. 
We also include negative-binomial expert models because they are widely used for overdispersed count outcomes, particularly in gene-expression applications \citep{love2014moderated}. 

Simulation~1, presented in detail below, evaluates estimation accuracy, the coverage of the simultaneous confidence bands and its dependence on sample size, classification performance relative to a constant-coefficient mixture-of-experts model, and the null distribution of the generalized likelihood ratio statistic. Simulations~2--4 extend these assessments to other response distributions and larger numbers of components, and are summarized in Supplementary Material~S7.

To evaluate the accuracy of the estimated functions, we employ the root average squared error (RASE). For a given coefficient function $\beta_p(\cdot)$, the RASE is defined as
$
    \text{RASE}_{\beta_p}
    = \sqrt{N^{-1}\sum_{j=1}^N \bigl(\hat{\beta}_p(u_j) - \beta_p(u_j)\bigr)^2},
$
where $\beta_p(u_j)$ denotes the true underlying coefficient function evaluated at $u_j$ and $N$ is the number of local models, as defined in Section~\ref{subsec_est}. The same criterion is evaluated for the components of $\boldsymbol{\alpha}(\cdot)$ and $\delta(\cdot)$, respectively.

\subsection{Simulation 1: two-component Gaussian expert model}\label{example1}
Consider a two-component mixture of varying-coefficient models obtained by specifying Model (\ref{main_model}) with $C = 2$.  We first generate covariates $X$ and $Z$ from the standard normal distribution and draw $u$ from the uniform distribution $U(0,1)$. To generate $Y$, we specify $\phi\{\cdot\}$ as a Gaussian distribution density function, $g(\cdot)$ as an expit function. We first consider \textbf{Coefficient Setting 1}, where the coefficient functions are specified as follows:
\begin{equation}\label{sim1_set}
\begin{aligned}
    \beta_0(u)   &= -0.4 + u, 
    & \beta_1(u)   &= 0.9-1.2u, \\[0.5em]
    \alpha_{10}(u) &= -0.5 + 0.6\cos(2\pi u), 
    & \alpha_{11}(u) &= 1 + 0.6\sin(2\pi u), \\[0.5em]
    \alpha_{20}(u) &= 0.5 + 0.6\cos(2\pi u), 
    & \alpha_{21}(u) &= 2 + 0.6\sin(2\pi u), \\[0.5em]
    \delta_{1}(u)  &= 0.85 + 0.35\cos(2\pi u), 
    & \delta_{2}(u)  &= 1.85 + 0.35\cos(2\pi u).
\end{aligned}
\end{equation}
The sample size is fixed at $n=500$, and the simulations are repeated 200 times.

We implement the VCMoE method as described in Section \ref{subsec_est} on the simulated data, where the kernel function $K(t)$ in the estimation is chosen as the Epanechnikov kernel $K(t) = 0.75(1-t^2)_{+}$. Following  the likelihood cross-validation criterion described in Section~\ref{sec:bandwidth_selection}, the selected optimal bandwidth is $h = 0.21$. To assess the performance of the method under this choice and its sensitivity of $h$, we additionally consider two bandwidths: $h=0.18$ and $h=0.24$, respectively, corresponding to values below and above the optimal choice. The performance is evaluated by RASE.

To evaluate the sensitivity of the proposed model to the specification of coefficient functions, we additionally consider \textbf{Coefficient Setting~2}, in which we set $\beta_1(u)=0.5$, $\alpha_{11}(u)=1$, $\alpha_{21}(u)=2$, $\delta_{1}(u)=0.35+0.35\cos(2\pi u)$, and $\delta_{2}(u)=1.85+0.35\cos(2\pi u)$ as defined in \eqref{sim1_set}, while keeping all remaining coefficient functions identical to those in \eqref{sim1_set}. Under \textbf{Coefficient Setting~2}, the optimal bandwidth selected via likelihood cross-validation is $h=0.29$; for comparability with \textbf{Coefficient Setting~1}, we further examine two additional bandwidth values, $h=0.26$ and $h=0.32$. Notably, in \textbf{Coefficient Setting~2}, all coefficients are estimated as fully functional forms, disregarding any prior knowledge that certain coefficients are in fact constants. This design enables us to examine the estimation behavior of truly functional coefficients when certain other coefficients are, in fact, constant.

\begin{figure}[ht]
    \centering
    \includegraphics[width=\textwidth,height=0.83\textheight,keepaspectratio]{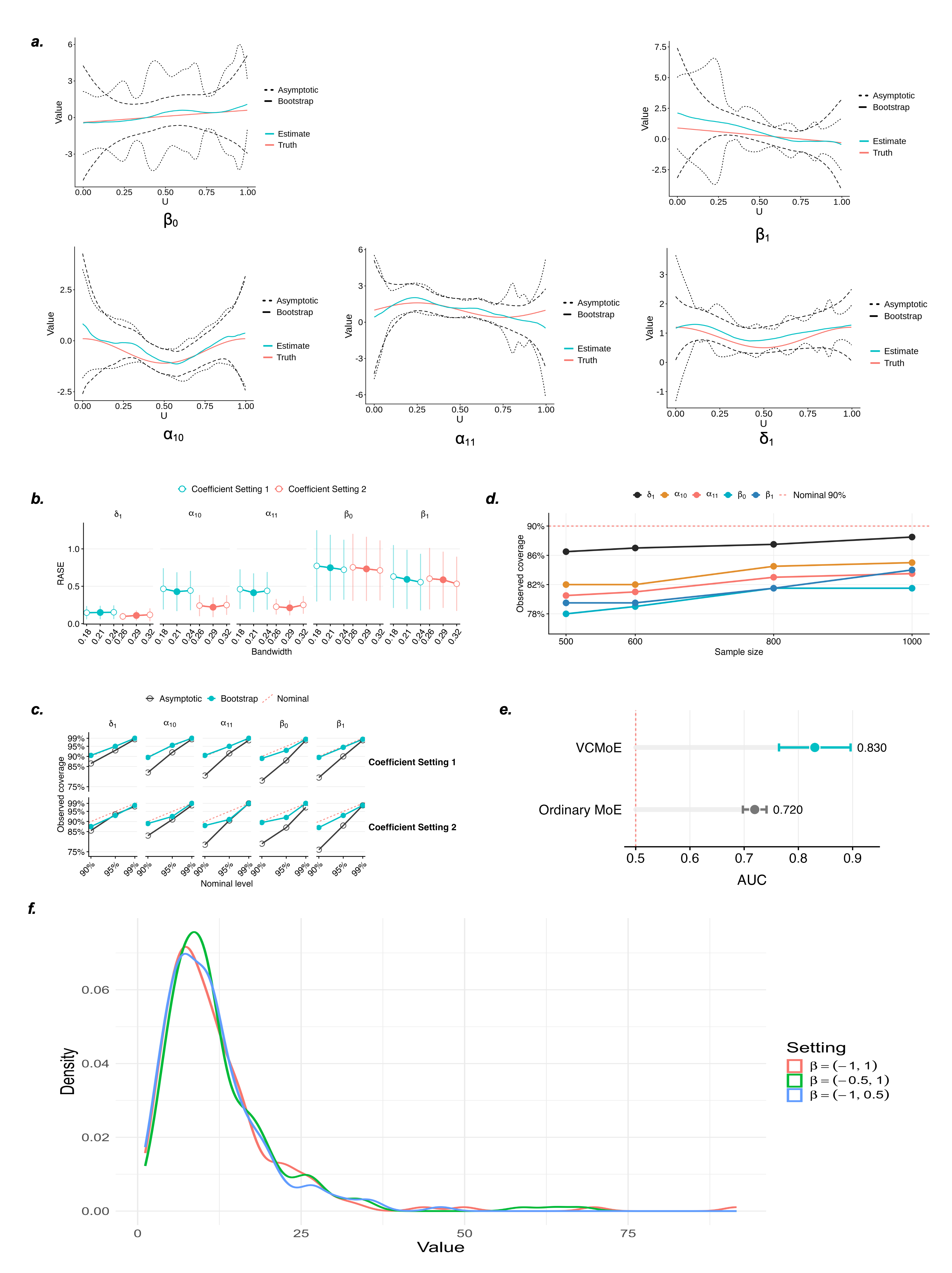}
    \caption{
Summary of Simulation~1 for the two-component Gaussian expert model.
(a) Estimated coefficient functions and the corresponding true functions under Coefficient Setting~1 with $n=500$ (Sample id \#1), together with asymptotic and bootstrap simultaneous confidence bands.
(b) Mean RASE values across 200 replications under different bandwidths for Coefficient Settings~1 and~2; vertical ranges indicate mean $\pm$ SD and filled points mark the bandwidth selected by likelihood cross-validation.
(c) Empirical coverage rates of simultaneous confidence bands at nominal levels $90\%$, $95\%$, and $99\%$, comparing the asymptotic and bootstrap approaches; the dashed orange diagonal denotes nominal coverage.
(d) Empirical coverage rates of the asymptotic $90\%$ confidence bands under increasing sample sizes for Coefficient Setting~1; the dashed orange horizontal line denotes nominal $90\%$ coverage.
(e) Mean AUC values for VCMoE and ordinary MoE, with horizontal error bars indicating SD across replications.
(f) Empirical densities of the generalized likelihood ratio test statistic $\lambda_n$ under the null hypothesis for three constant values of $\boldsymbol{\beta}$.
}
\label{fig:sim1_summary}
\end{figure}

The mean and standard deviation of RASEs for \textbf{Coefficient Setting 1} and \textbf{Coefficient Setting 2}, computed over 200 replications, are reported in Figure~\ref{fig:sim1_summary}(b). The results show that not all RASEs attain their minimum at the selected optimal bandwidth, suggesting that the coefficient functions $\boldsymbol{\beta}(u)$, $\boldsymbol{\alpha}(u)$, and $\boldsymbol{\delta}(u)$ may possess different degrees of smoothness. We also observe that the RASEs for the coefficient estimates in the gating function, i.e., $\boldsymbol{\beta}(\cdot)$, are larger than those for the coefficients in the expert models, i.e., $\boldsymbol{\alpha}(\cdot)$ and $\boldsymbol{\delta}(\cdot)$. This result is expected, as the gating function involves latent parameters, which are inherently subject to higher estimation uncertainty. Moreover, the RASE values under \textbf{Coefficient Setting~2} exhibit patterns similar to those observed under \textbf{Coefficient Setting~1}, with slightly smaller overall magnitudes. These results indicate that the proposed model remains robust to misspecification arising from treating truly constant coefficients as functional.

Next, we construct simultaneous confidence bands described in Section \ref{subsec_cb} for the coefficient functions using both the asymptotic distribution approach (Section \ref{cb_asy}) and the bootstrap approach (Section \ref{cb_boot}) in both coefficient settings. To reduce the impact of bias, we adopt an undersmoothing strategy by selecting a smaller bandwidth $h=0.18$ for \textbf{Coefficient Setting 1} and $h=0.26$ for \textbf{Coefficient Setting 2}. This is a common practice for constructing simultaneous confidence bands, where the bandwidth is often taken to be \(80\%\)–\(90\%\) of the optimal choice, in varying-coefficient models (see \cite{fan2000simultaneous, zhang2010simultaneous}). We then compute the coverage probabilities of the resulting confidence bands at the nominal confidence levels of $90\%$, $95\%$, and $99\%$, respectively, with results summarized in Figure~\ref{fig:sim1_summary}(c). It is evident that the bootstrap approach outperforms the asymptotic-distribution-based approach in both settings. An illustrative example of the estimated coefficient function, together with its simultaneous confidence bands obtained from the asymptotic and bootstrap approaches for both settings, is presented in Figure~\ref{fig:sim1_summary}(a) and Figure S1 in the Supplementary Material, respectively, where we observe signs of instability in the covariance matrix estimation.
This pattern is consistent with the observation in \citet{yang2024estimation}, suggesting that the bootstrap approach can be more stable in finite-sample settings. We return to this issue in greater detail in Simulation~3. 

 To examine the effect of sample size on the coverage rate of the asymptotic approach, we repeat the simulation studies in \textbf{Coefficient Setting 1} but increase the sample sizes to 600, 800, and 1000, respectively. In this simulation study, we focus on the 90\% confidence level where severe undercoverage is observed. The results, summarized in Figure~\ref{fig:sim1_summary}(d), indicate that as sample size increases, the asymptotic confidence bands achieve improved coverage rates.

We further evaluate the classification performance of the proposed VCMoE model against a conventional mixture-of-experts (MoE) model, in which the coefficients are assumed to be constant with respect to the index variable. For each Monte Carlo replication under coefficient setting 1, we compute the area under the receiver operating characteristic curve (AUC). Figure~\ref{fig:sim1_summary}(e) shows that VCMoE achieves substantially better classification performance than the conventional MoE model, achieving a mean AUC of 0.830 compared with 0.720 for the ordinary MoE model.

Finally, we investigate a Wilks phenomenon when applying the generalized likelihood ratio test (GLRT) statistic (as described in Section \ref{glrt}) for testing $H_0: \boldsymbol{\beta}(\cdot)=\boldsymbol{\beta}$. We focus on the parameter $\boldsymbol{\beta}$, the parameter that presents in the mixing proportion function, since estimation of non-constant mixing proportions is the key innovation in this article. The data-generating process is the same as in the previous setting, except that $\boldsymbol{\beta} (u)$ in (\ref{sim1_set}) is now taken to be a constant vector. Therefore, we consider the \textbf{Coefficient Setting 3}, where we set the true values of $\boldsymbol{\beta}$ to be $(-1,1),\, (-0.5,1),$ and $(-1,0.5)$, respectively. 
The hybrid estimation method described in Section~\ref{est_cons} is used to compute the log-likelihood $\ell_n(H_0)$ under the null hypothesis, and the method described in Section~\ref{m_em} is used to estimate the log-likelihood $\ell_n(H_a)$ under the alternative hypothesis. 
For each specification of $\boldsymbol{\beta}$, the simulation is repeated 200 times to approximate the distribution of the test statistic $\lambda_n$. 
This empirical distribution serves as a proxy for the true unconditional distribution of the test statistic. 
The three resulting density curves, shown in Figure~\ref{fig:sim1_summary}(f), are nearly identical. 
This finding is consistent with Theorem~\ref{theorem_glrt}, which establishes that the asymptotic distribution of $\lambda_n$ under the null hypothesis is independent of the true values of the unknown constant coefficients and other nuisance parameters.

\subsection{Simulations 2--4: Additional expert models and response distributions}

We conduct three further simulation studies, summarized here with full specifications, tables, and figures in Supplementary Material~S7. Simulation~2 replaces the Gaussian experts with binomial logistic experts; estimation accuracy, the bootstrap advantage in confidence-band coverage, and the Wilks phenomenon all mirror Simulation~1, with the asymptotic bands becoming more stable as the binomial total count increases. Simulation~3 enlarges the mixture to three Gaussian components under a softmax gating function, reproduces the same estimation and coverage patterns, and shows that the asymptotic-band instability noted in Simulation~1 recurs under Gaussian experts but is absent under binomial experts. Simulation~4 is the most demanding and the most relevant to our application: it fits five-component models under Gaussian, binomial, and negative-binomial experts sharing a common softmax gating structure, with the negative-binomial experts carrying a library-size offset as is standard for gene-expression counts \citep{zhang2019probabilistic}. Across all three response families the method attains accurate estimation, and posterior assignment is recovered well for the well-separated components, with the expected loss of accuracy for the overlapping middle components.

\section{Application to mouse embryonic snRNA-seq data}\label{sec:Application}
In this section, we use VCMoE to analyze single-nucleus RNA sequencing (snRNA-seq) data obtained during embryonic development of the house mouse. Our primary objective is to characterize how the associations between selected genes, expressed in neurons, may evolve across embryonic days of brain cortex development. What sets VCMoE apart from existing methods is that it separates a genuine change in a within-subpopulation association from a concurrent shift in subpopulation composition, so the former can be inferred without being confounded by the latter. Our central finding on the developing cortex is that the repression of \textit{Bcl11b} by \textit{Satb2} in upper-layer neurons is not yet in place when those neurons first appear and becomes established over the course of their differentiation, a dynamic that a constant-coefficient mixture-of-experts model averages into a single intermediate value. Because the cell-type annotations are withheld from the fitting throughout, that reading rests on the fitted components corresponding to the neuronal subtypes they are taken to represent, and we show that they do.

\subsection{Data and model specification}\label{subsec:app_data}

The dynamic developmental process in the mouse brain cortex reflects changes in two major cortical neuron subtypes, deep-layer and upper-layer neurons, whose relative abundance and cellular composition change over embryonic development. Deep-layer neurons develop earlier, and their axons establish early trajectories that form the backbone of later-developing cortical circuits. Upper-layer neurons develop later, and often extend their axons along the pioneer trajectories laid by the deep-layer neurons. Their development is guided by molecular cues from the deep-layer neurons \citep{toma2014timing}. Therefore, gene--gene associations are expected to change over embryonic time, while the relative composition of deep-layer and upper-layer neurons is also shifting. This situation motivates our use of the VCMoE model to capture these dynamic, subtype-driven patterns, by modeling these two subtypes of neurons as two latent classes within the framework.  We accordingly set $C=2$, matching the two neuronal subtypes that the scientific question concerns. Section~\ref{subsec:app_c3} reports a sensitivity analysis under a third latent class, and Section~\ref{sec:Discussion} discusses the general problem of an unknown $C$.

We obtained snRNA-seq data from 12.4~million nuclei extracted from 83~mouse embryos, where the embryos were sampled at 2-6 hour intervals in prenatal development between gastrulation (approximately embryonic day 8) and birth \citep{qiu2024single}. The cells were previously annotated into hundreds of cell types in order to investigate developmental patterns of many embryonic structures in the mouse.

We restricted our attention to the deep-layer and upper-layer neuronal subtypes, between embryonic day~14 (E14) and embryonic day~18.75 (E18.75), where the latter is the final embryonic stage before birth, and the former is approximately when upper-layer neurons begin to be generated \citep{toma2014timing}, so that E14 is the earliest stage at which both subtypes are represented. At each of the 20 developmental time points, we sampled 2,000 neurons, using stratified sampling to preserve the cell-type composition, giving 40,000 neurons in total. Although the cell types had been previously assigned, we intentionally exclude this information from our modeling steps and treat the cell-type structure as latent. The annotations are not direct observations of cell identity: they were obtained by clustering cells on their transcriptome-wide expression profiles and then assigning an identity to each cluster from combinations of marker genes \citep{qiu2024single}, and so carry their own error. Conditioning on them would import that error into the association analysis without accounting for it, so that estimated labels would be treated as known and the resulting uncertainty understated. Holding them out instead leaves them available as a separate characterization of the same cells, against which the fitted components are checked in Section~\ref{subsec:app_classification}.

As our response variable, we choose the expression level of \textit{Bcl11b}, a gene considered to be a canonical identifier of deep-layer neurons, denoted as $Y^{\text{Bcl11b}}$. We are particularly interested in the association between the expression levels of \textit{Bcl11b} and \textit{Satb2}, because previous studies have demonstrated that \textit{Satb2} acts as a negative regulator of \textit{Bcl11b} during cortical development \citep{britanova2008satb2}. To also validate model performance in a situation where no association is expected (i.e. a negative control), we also investigate the association between the expression levels of \textit{Ywhaz}, a gene whose expression is expected to be approximately constant over developmental time. \textit{Ywhaz} is a known housekeeping gene \citep{shaydurov2018analysis}.  We further include two biologically relevant transcription factors, \textit{Fezf2} and \textit{Pou3f2}. \textit{Fezf2} is involved in specifying deep-layer projection-neuron identity, with \textit{Bcl11b} serving as an important downstream effector in this regulatory program \citep{chen2008fezf2}. In contrast, \textit{Pou3f2} is associated with the upper-layer cortical neuron program \citep{dominguez2013pou}. Therefore, the covariate vector in the expert model is specified as \(\boldsymbol{z} = (z^{\text{Satb2}}, z^{\text{Ywhaz}},z^{\text{Fezf2}},z^{\text{Pou3f2}})^\top\), where each component denotes the expression level of the corresponding gene. As the latent cell types, upper- and deep-layer neurons are characterized by their marker genes \textit{Satb2} and \textit{Ntng1}, respectively \citep{fujita2020netrin}. Accordingly, we consider that the covariates entering the gating functions are given by \(\boldsymbol{x} = (x^{\text{Satb2}}, x^{\text{Ntng1}})^\top\), where \(x^{\text{Satb2}}\) and \(x^{\text{Ntng1}}\) denote the expression levels of \textit{Satb2} and \textit{Ntng1}, respectively. A descriptive summary of the expression levels of these genes of interest across the two neuronal cell types is presented in Figure~\ref{fig_real_dataset}(a). It can be seen that the expression of \textit{Bcl11b} is substantially higher in deep-layer neurons than in upper-layer neurons, whereas \textit{Satb2} exhibits higher expression in upper-layer neurons and comparatively low expression in deep-layer neurons.

Then, we use model (\ref{main_model}) to carry out the analysis. Specifically, we model \(Y_i^{\textit{Bcl11b}}\) using a two-component negative-binomial mixture, a standard choice for overdispersed gene-expression count data \citep{zhang2019probabilistic}. The conditional probability mass function is
\begin{equation}
    \begin{aligned}
        f\!\left(Y_i^{\textit{Bcl11b}}\right)
        =
        &\; \pi(u_i;\boldsymbol{x}_i)\,
        \phi_1\!\left(Y_i^{\textit{Bcl11b}}; \eta_{1i}, \delta_1(u_i)\right)
        \\
        &+
        \{1-\pi(u_i;\boldsymbol{x}_i)\}\,
        \phi_2\!\left(Y_i^{\textit{Bcl11b}}; \eta_{2i}, \delta_2(u_i)\right),
    \end{aligned}
\end{equation}
where \(\phi_c(\cdot;\eta_{ci},\delta_c(u_i))\) denotes the negative-binomial probability mass function for component \(c\), with mean \(\eta_{ci}\) and dispersion parameter \(\delta_c(u_i)\). The component-specific mean is modeled through the log link \(\log \eta_{ci} = \log s_i + \boldsymbol{z}_i^\top \boldsymbol{\alpha}_c(u_i)\), for $c=1,2$, where $c=1$ and $c=2$ correspond to upper- and deep-layer neurons, respectively. The term $\log s_i$ is included as a library-size offset to account for variation in sequencing depth \citep{zhang2019probabilistic}. The covariate vector is \(\boldsymbol{z}_i = (1, z_i^{\textit{Satb2}}, z_i^{\textit{Ywhaz}}, z_i^{\textit{Fezf2}}, z_i^{\textit{Pou3f2}})^\top\), and the corresponding component-specific varying-coefficient vector is \(\boldsymbol{\alpha}_c(u_i) = (\alpha_{c0}^{\mathrm{int}}(u_i), \alpha_{c1}^{\textit{Satb2}}(u_i), \alpha_{c2}^{\textit{Ywhaz}}(u_i), \alpha_{c3}^{\textit{Fezf2}}(u_i), \alpha_{c4}^{\textit{Pou3f2}}(u_i))^\top\). Furthermore, \(\pi(u_i; \boldsymbol{x}_i) = \operatorname{expit}\{\beta_0^{\mathrm{int}}(u_i) + x_i^{\mathrm{Satb2}} \beta_1^{\mathrm{Satb2}}(u_i) + x_i^{\mathrm{Ntng1}} \beta_2^{\mathrm{Ntng1}}(u_i)\}\) denotes the conditional probability that cell \(i\) belongs to the upper-layer neuron.  All covariates are preprocessed using standard library-size normalization followed by log transformation \citep{stuart2019comprehensive}. For model fitting, we employ the Epanechnikov kernel for its asymptotic efficiency \citep{wand1994kernel}.
The developmental time points are rescaled to the interval $[0,1]$ based on their original temporal scale, and the bandwidth is chosen to be 0.22 by the likelihood cross-validation criterion. 

\subsection{Dynamic gene--gene associations}\label{subsec:app_dynamics}

\begin{figure}[ht]
    \centering
    \includegraphics[width=\textwidth,height=0.83\textheight,keepaspectratio]{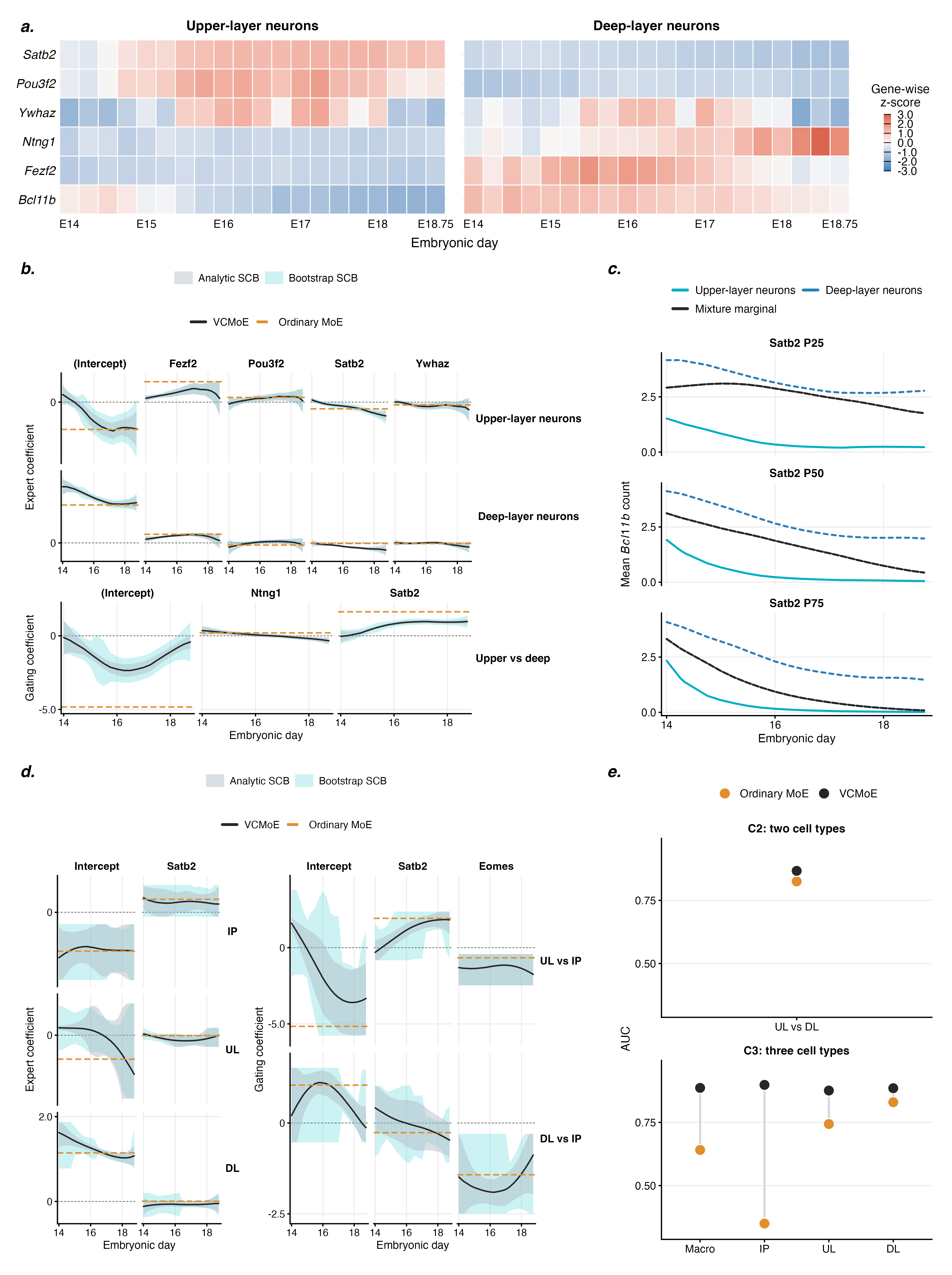}
   \caption{
Summary of the application to mouse embryonic single-nucleus RNA-sequencing data.
(a) Heatmap showing gene-expression patterns in upper- and deep-layer neurons.
(b) VCMoE estimates of the coefficient functions, with simultaneous confidence bands, together with the corresponding constant-coefficient estimates from the ordinary MoE model in the two-cell-type analysis.
(c) Estimated effect of \textit{Satb2} expression on \textit{Bcl11b} at selected empirical percentiles of \textit{Satb2}, with all other covariates fixed.
(d) Sensitivity analysis under the three-cell-type specification, showing the VCMoE coefficient-function estimates with simultaneous confidence bands and the corresponding constant-coefficient estimates from the ordinary MoE model.
(e) Concordance between the fitted components and the withheld cell-type annotations, reported using the AUC for the two-cell-type analysis and the macro-AUC and component-specific AUCs for the three-cell-type analysis.
}

\label{fig_real_dataset}
\end{figure}

As shown in Figure~\ref{fig_real_dataset}(b), the estimated \textit{Satb2} coefficient function in upper-layer neurons, $\hat{\alpha}_{11}^{\textit{Satb2}}(\cdot)$, is not distinguishable from zero at E14, the stage at which these neurons first appear, and decreases steadily thereafter, with its bootstrap-based simultaneous confidence band lying entirely below zero from E15.75 onward. The repression of \textit{Bcl11b} by \textit{Satb2} is therefore not yet in place when upper-layer neurons are born, and becomes established as they differentiate. This ordering is consistent with the role of \textit{Satb2} as a postmitotic determinant of upper-layer identity \citep{britanova2008satb2}, and with the two genes being co-expressed at the earliest stages \citep{yang2024spatial}. In deep-layer neurons, by contrast, $\hat{\alpha}_{21}^{\textit{Satb2}}(\cdot)$ stays close to zero across the window, with only a slight decline at the latest stages, indicating that \textit{Satb2} acts on \textit{Bcl11b} specifically within the upper-layer program. The ordinary MoE model, constrained to a single constant, reports an intermediate negative coefficient throughout (Figure~\ref{fig_real_dataset}(b)). It therefore overstates the repression at E14, understates it at E18.75, and offers no representation of the onset itself: the stage at which the repression becomes established cannot be recovered from a single constant, whatever its value.

The simultaneous confidence bands provide further evidence that the \textit{Satb2}--\textit{Bcl11b} associations vary over development, as no single constant function lies entirely within either band over the developmental window. Consistently, generalized likelihood ratio tests reject the null hypotheses that $\alpha_{11}^{\textit{Satb2}}(\cdot)$ and $\alpha_{21}^{\textit{Satb2}}(\cdot)$ are constant, with both $p$-values below $10^{-16}$. The deep-layer coefficient varies detectably but stays within a narrow range close to zero, whereas the upper-layer coefficient moves from indistinguishable from zero to clearly negative. Figure~\ref{fig_real_dataset}(c) also presents the estimated \textit{Bcl11b} trajectories at selected empirical quantiles of \textit{Satb2}, with the remaining covariates fixed at their medians. The results identify E14--E16 as the principal transition period, with the steepest decline occurring between E14 and E15. During this interval, upper-layer neurons with high \textit{Satb2} expression rapidly shift toward a \textit{Bcl11b}-low state. This timing is consistent with the established period of upper-layer neurogenesis and postmitotic projection-neuron specification \citep{toma2014timing,britanova2008satb2}.

The remaining covariates separate into effects that are stable over development and effects that vary. The coefficient functions for \textit{Ywhaz} remain approximately constant throughout the developmental window, and their simultaneous confidence bands contain a constant function over the entire domain. Generalized likelihood ratio tests of constant effects yield $p$-values of 0.993 and 0.997 for the two neuronal components, respectively, providing no evidence of temporal variation. This result is consistent with the established role of \textit{Ywhaz} as a housekeeping gene. By contrast, \textit{Fezf2} exhibits a predominantly positive association with \textit{Bcl11b}, consistent with the role of \textit{Bcl11b} as a downstream effector of the \textit{Fezf2} regulatory program. Although \textit{Pou3f2} is also associated with upper-layer neuronal programs, its conditional association with \textit{Bcl11b} differs markedly from that of \textit{Satb2}. After adjustment for the other covariates, \textit{Pou3f2} is positively associated with \textit{Bcl11b} during part of the developmental window, consistent with previous observations \citep{dominguez2013pou}.

The gating estimates show that the composition following which the expert coefficients are estimated is itself shifting. The estimated gating coefficient $\hat{\beta}_1^{\textit{Satb2}}(\cdot)$ is positive with an increasing trajectory, consistent with \textit{Satb2} being a characteristic marker of upper-layer neurons and with the association between its expression and upper-layer identity strengthening over development, whereas $\hat{\beta}_2^{\textit{Ntng1}}(\cdot)$ becomes negative during the later stages, consistent with the preferential expression of \textit{Ntng1} in deep-layer neurons. Both gating coefficients therefore vary over the same window in which the \textit{Satb2} repression becomes established, so a model holding either the mixing proportions or the expert coefficients constant would attribute part of one change to the other.

\subsection{Correspondence with the annotated cell types}\label{subsec:app_classification}

To verify that the fitted components can be identified with the two neuronal subtypes whose coefficients are interpreted in Section~\ref{subsec:app_dynamics}, we compare the posterior component probabilities with the withheld annotations in a separate test dataset drawn by the same sampling procedure as the training data. The two characterizations share information without being equivalent. The identity assigned to each annotated cluster was read from a combination of marker genes (Section~\ref{subsec:app_data}), two of which, \textit{Satb2} and \textit{Ntng1}, also enter the gating; a close correspondence is to that extent expected. The annotated partition itself, however, derives from transcriptome-wide clustering, whereas the model sees only the genes specified in $\boldsymbol{x}$ and $\boldsymbol{z}$. The comparison therefore indicates whether the components can be identified with the annotated subtypes, not whether the model discovers those subtypes on its own.

VCMoE attains an AUC of 0.867 and the ordinary MoE model attains a slightly lower value, 0.825 (Figure~\ref{fig_real_dataset}(e)). At $C=2$ the two subtypes are separated well enough that both models locate them. The components of both models align with the same annotated subtypes, so the constant coefficient and the coefficient function contrasted in Section~\ref{subsec:app_dynamics}, and the models compared in Section~\ref{subsec:app_modelcomp}, describe the same two subpopulations rather than different partitions of the data.

\subsection{Model comparison}\label{subsec:app_modelcomp}

To separate a change in the within-subpopulation association from the concurrent shift in subpopulation composition, we compare three nested models on the same cells, response, and covariate pools: (i) the ordinary MoE model, in which both the gating and the expert coefficients are constant in $u$, already shown in Figure~\ref{fig_real_dataset}(b); (ii) a model in which the expert coefficients vary with $u$ while the gating coefficients are held constant (static gating); and (iii) the proposed model, in which both vary with $u$ (VCMoE). The ordinary MoE model omits two components at once, and thus a comparison between it and VCMoE does not indicate which of the two accounts for the estimated onset of \textit{Satb2} repression. Therefore, the static-gating model is included to separate them, as comparing it with the ordinary MoE model quantifies the contribution of letting the expert coefficients vary with $u$, and comparing VCMoE with it quantifies the contribution of letting the gating vary with $u$.

Table~\ref{tab:modelcomp} reports the time-stratified cross-validated predictive log-likelihood and the AUC against the held-out annotations for each model. Model performance improves progressively from the ordinary MoE model to the static-gating model and then to VCMoE, providing further evidence that allowing the coefficients to vary with the indexing variable is important for adequately capturing the underlying dynamics.

\subsection{Sensitivity to a third latent cell type}\label{subsec:app_c3}

As a sensitivity analysis, we extend the model by including intermediate neuronal progenitors as a third latent cell type such that $C=3$. We retain the same sampling scheme and general model specification, while adding a third mixture component and including \textit{Eomes}, a marker of intermediate neuronal progenitors, as an additional gating covariate \citep{qiu2024single}. Because intermediate progenitors give rise to both upper- and deep-layer neurons and become increasingly rare over the developmental window, they constitute a particularly challenging subpopulation to identify. Nevertheless, as shown in Figure~\ref{fig_real_dataset}(d), VCMoE recovers the same general \textit{Satb2}--\textit{Bcl11b} association patterns within the upper- and deep-layer neuronal components. Notably, the estimated association is positive within the intermediate-progenitor component, further illustrating the cell-type-specific nature of this relationship. This finding is compatible with early co-expression of \textit{Satb2} and \textit{Bcl11b} before lineage-specific repression becomes established during differentiation into upper-layer neurons. The estimated gating effects of \textit{Eomes} for the neuronal components remain negative throughout the developmental window, consistent with its preferential expression in intermediate neuronal progenitors.

The correspondence with the annotations, close for both models at $C=2$, differs between them at $C=3$. VCMoE attains a macro-AUC of 0.887 and the ordinary MoE model attains 0.641, a gap driven largely by the progenitors, for which the component-specific AUC is 0.350 under the ordinary MoE model and 0.899 under VCMoE (Figure~\ref{fig_real_dataset}(e)). This may be attributed to the transience of the progenitor state: a model whose coefficients are constant in $u$ cannot represent a component that is present early and largely absent by E18.75, and its third component does not align with the progenitors.

\section{Discussion}\label{sec:Discussion}
In this article, we propose a Varying-coefficient Mixture-of-Experts (VCMoE) model that extends the classical Mixture-of-Experts framework by allowing all coefficients in both the gating and expert components to vary smoothly with an indexing variable. For theoretical development, we focus on the two-component case for clarity of exposition, while numerical studies consider two-, three-, and five-component settings across a range of response types. We establish identifiability and asymptotic convergence, develop a tailored EM algorithm for estimation, and derive inferential procedures for uncertainty quantification and hypothesis testing. We apply the method to a mouse embryonic snRNA-seq atlas covering the window over which cortical projection neurons are specified.

The estimated \textit{Satb2} coefficient function in upper-layer neurons is not distinguishable from zero at E14, when these neurons first appear, and its simultaneous confidence band lies entirely below zero from E15.75 onward. The repression of \textit{Bcl11b} is therefore not yet in place at the time upper-layer neurons are born, and becomes established as they differentiate. The corresponding coefficient in deep-layer neurons stays close to zero across the window. The gating coefficients change over the same window, so the composition of the two subpopulations shifts while the within-subpopulation associations are themselves changing. A model holding all coefficients constant reports a single intermediate negative value, from which the timing of the onset cannot be recovered.

That \textit{Satb2} represses \textit{Bcl11b}, and that the two genes are co-expressed early in development, are both established. Existing reports, however, describe expression at particular developmental stages, in cells whose identity is fixed in advance. The analysis in Section~\ref{sec:Application} instead tests whether the association between the two genes changes over developmental time, with uncertainty quantification and without treating the cell-type labels as known. The estimated onset can then be read against what is already understood about cortical development.


One limitation of our model is that we assume that the number of latent classes is known. In practice, this assumption may not hold, particularly in settings where prior domain knowledge is unavailable. Possible extensions include a Bayesian nonparametric formulation, such as a Dirichlet process mixture, or an EM-type test for determining the order of the mixture \citep{li2010testing}. Extending such order-selection procedures to VCMoE represents an important direction for future research. Another limitation is that the current formulation considers a one-dimensional indexing variable. Extending the framework to multidimensional indices would broaden its applicability to settings such as spatial transcriptomics, where the underlying heterogeneity may vary over a two-dimensional spatial domain \citep{zhao2025distect}.

\section{Data Availability Statement}\label{data-availability-statement}

The snRNA-seq dataset analyzed in this study is publicly available at \url{https://omg.gs.washington.edu/} and via \href{https://cellxgene.cziscience.com/collections/45d5d2c3-bc28-4814-aed6-0bb6f0e11c82}{CELLxGENE}. The R package implementing the proposed method, \texttt{VCMoE}, is available from CRAN at \url{https://CRAN.R-project.org/package=VCMoE}. A GitHub repository for ongoing development, documentation, and issue tracking is maintained at \url{https://github.com/qc-zhao/VCMoE}.

\section{Acknowledgments}
Qihuang Zhang was supported by the Natural Sciences and Engineering Research Council of Canada (NSERC) and the Canadian Statistical Sciences Institute (CANSSI) Quebec, and his research was undertaken, in part, thanks to funding from the FRQ-Sant\'e Program. Zhang is a Fonds de recherche du Qu\'ebec Research Scholar (Junior~1).
      Celia M.T.\ Greenwood was supported by the Arthritis Society Canada Strategic Operating Grant SOG-23-0261.

\section{Supplementary material}
\label{SM}
 The Supplementary Material presents theoretical results on identifiability and the consistency of global estimators, supplementary figures, additional simulation results, and all proofs.

\section{Competing interests}
The authors declare that they have no conflicts of interest.



\bibliographystyle{abbrvnat}
\bibliography{reference}
\clearpage
\begin{table}[tbp]
\centering
\caption{Model comparison on the mouse embryonic snRNA-seq data. All three models use covariate-dependent gating and the same cells, response, and covariate pools. The ordinary MoE model is fitted by maximum likelihood on the pooled cells, developmental time playing no part; the static-gating and VCMoE fits use the same kernel and the same bandwidth selection procedure. Predictive log-likelihood is time-stratified and cross-validated; AUC is computed against the held-out annotations.}
\label{tab:modelcomp}
\small
\begin{tabular}{@{}lccrc@{}}
\hline
Model & Experts vary & Gating varies & Predictive & AUC \\
      & in $u$       & in $u$        & log-lik.   &     \\
\hline
Ordinary MoE    & no  & no  & -66,802.89 & 0.825 \\
Static gating   & yes & no  & -64,216.22 & 0.831 \\
VCMoE           & yes & yes & -63,921.07 & 0.867 \\
\hline
\end{tabular}
\end{table}

\end{document}